\documentclass[twocolumn,10pt]{asme2ej}

\usepackage{array}
\usepackage{url}
\usepackage{float}
\usepackage{amsmath}
\usepackage{amsfonts}
\usepackage{amssymb}
\usepackage{mathtools}
\usepackage{slashed}
\DeclareMathOperator*{\argmax}{arg\,max}
\DeclareMathOperator*{\argmin}{arg\,min}

\DeclarePairedDelimiter\floor{\lfloor}{\rfloor}
\usepackage{epsfig} 

\title{Solving the Optimal Trading Trajectory Problem Using Simulated Bifurcation}

\author{Kyle Steinhauer 
    \affiliation{
	Scientist at AlpacaJapan\\
	AlpacaJapan Co., Ltd.\\
	Email: kyle@alpaca.ai\\
    }	
    \and 
	\textbf{Takahisa Fukadai}
    \affiliation{
	Scientist at AlpacaJapan\\
	AlpacaJapan Co., Ltd.\\
	Email: taka@alpaca.ai\\
    }
}

\author{Sho Yoshida 
    \affiliation{
	Chief Data Scientist at AlpacaJapan\\
	AlpacaJapan Co., Ltd.\\
	Email: yoshiso@alpaca.ai
    }	
}

\begin{document}

\maketitle   
\section*{Abstract}
\begin{abstract}
{\it 
We use an optimization procedure based on simulated bifurcation (SB) to solve the integer portfolio and trading trajectory problem with an unprecedented computational speed. The underlying algorithm is based on a classical description of quantum adiabatic evolutions of a network of non-linearly interacting oscillators. This formulation has already proven to beat state of the art computation times for other NP-hard problems and is expected to show similar performance for certain portfolio optimization problems. Inspired by such we apply the SB approach to the portfolio integer optimization problem with quantity constraints and trading activities. We show first numerical results for portfolios of up to 1000 assets, which already confirm the power of the SB algorithm for its novel use-case as a portfolio and trading trajectory optimizer.
}
\end{abstract}

\section{Introduction}
The trading trajectory problem can be described as the problem to find the optimal set of portfolios and respective trading activities, which maximize the future expected return over a given time period while taking into account trading costs and expected risks. The optimal trajectory doesn't necessarily maximize the return at each time point, but  maximizes the future return over the entire time-period including all trading activities. We will follow the mean variance portfolio description \cite{markowitz1, markowitz2, markowitz3} in order to express portfolio values at each point in time and will add cost terms that account for the necessary rebalancing from one time step to the other.

There are many scenarios in asset management where trading and investment activities are constraint. Some common constraints lead to the problem of finding the optimal trading trajectory with only integer-valued solutions. This occurs for example in ETF block trades, where only a certain integer amount of a standard lot-size can be traded. The integer portfolio formulation (in its general form) belongs to the non-convex mixed integer quadratic problems and therefore falls into the class of NP-hard problems \cite{bienstock, lagrangianrelaxation}. This strongly complicates the search of the optimal trajectory and particularly effects the computation time with increasing system size. A further complication comes from quantity and cardinality constraints that are required in almost any real world application. A variety of studies \cite{intpf1,intpf2,intpf3,intpf4,intpf5} have been conducted to push the understanding and computational performance of the integer portfolio optimization with constraints. A branch and bound method \cite{exactintegersolution} for example, allowed to find the exact solution to a single time period portfolio optimization problem with up to 200 assets.

In the last couple of years we saw substantial progress in building quantum annealer systems and actual quantum computers, and the interest is rising to harness this technology for real world applications like the discussed mixed integer quadratic problems. An important step into this direction has been conducted by G. Rosenberg et al. \cite{dwaveportfolio}, which solved the integer trading trajectory problem on D-Wave's quantum annealer system. Very recent studies \cite{dwaveportfolio2} have shown continued interest and progress in this direction. Despite the increasing accessibility and power of those machines, the actual business applicability in finance is however still missing certain technological advancements.

H. Goto et al. however, have been studying quantum adiabatic evolution in detail, and inspired by such presented a recent formulation \cite{simulatedbifurcation} that can potentially bring quantum computer like speed for certain descriptions of NP-hard problems to classical computers. In their recent study\cite{simulatedbifurcation} they introduced a formulation of the Ising model using Kerr-nonlinear parametric oscillators, which is highly effective in solving the Ising optimization problem via its quantum adiabatic evolution through its bifurcation point. This simulated bifurcation (SB) algorithm is highly suitable for parallel computing and beats (around $10$ $\times$ faster) the current state of the art custom built machine to solve a fully connected 2000 spin problem \cite{isingmachine}. Since many NP-hard problems can be expressed in form of an Ising problem \cite{manynpproblems}, the simulated bifurcation has the potential to boost computational performance for a large range of those combinatorially difficult problems. The discovery of this highly promising formulation has also triggered larger institutions, like Toshiba for example, to construct dedicated simulated bifurcation machines (SBM) and already advertise potential real world use-cases, e.g. the detection of triangular arbitrage.

In this work we make use of this novel SB-formulation to solve the optimal portfolio and optimal trading trajectory problem. For this we will first map the integer portfolio problem to the respective Ising problem, recapitulate the mentioned SB-algorithm, and then use our implementation to generate first results of optimal portfolios and trajectories found by simulated bifurcation.

\section{Ising Problem and Finding the Optimal Solution}
The Ising problem \cite{isingmodel} can be formulated as the problem to find a ground-state (spin configuration) that minimizes the Ising energy defined by
\begin{equation}
\label{eq:isingenergy}
E = -\frac{1}{2} \sum_{i,j}J_{ij}s_i s_j +\sum_i h_i s_i,
\end{equation}
where $s_i \in \{-1, 1\}$ are commonly interpreted as spins pointing down or upwards. Note that both $J_{ij}$ and $h_{i}$ are $\in \mathbb{R}$ and the coupling among the spins is symmetric, such that $J_{ij} = J_{ji}$. The sum runs over all $N$ spin-variables in the system. This problem has been studied in connection to various fields of physics, mainly due to the spontaneous symmetry breaking that occurs in 2 or more dimensions. This is something observed in a real world ferromagnet but also the mechanism behind the mass generation in the Standard Model of particle physics \cite{higgs1, higgs2, higgs3}. For this work however, the more important point of interest is the combinatorial and computational aspect of the problem. Without any topological structure, the fully connected Ising problem falls into the class of NP-hard problems \cite{spinglass}. The number of possible states is $2^N$, where $N$ is the number of spins in the system. For decently sized systems it is impossible to deterministically find the ground-state in a reasonable amount of time on a classical computer by iterating through all combinations. Monte-Carlo algorithm like the Metropolis algorithms and many others, are successfully used to sample the different spin states and to approach the ground-state without generating all combinatorial possibilities. Note however, that for fully connected Ising systems (non-local actions), most of these algorithms show rather weak performance, are not suitable for parallel computing, or simply not usable.

The already mentioned work \cite{simulatedbifurcation} by H. Goto et al. describes the Ising system with Kerr-nonlinear parametric oscillators and proposes a new optimization algorithm that simulates adiabatic evolutions of classical nonlinear Hamiltonian systems. If the system is initialized in a proper vacuum-state, the so called simulated bifurcation (SB) algorithm simulates the adiabatic evolution of the system through its bifurcation point and ends up in a ground-state that minimizes the Ising energy. Using this formulation has shown to beat state of the art computation times for large fully connected systems. Note that for smaller systems, different heuristic approaches, like the so called Digital Annealer \cite{digitalannealer}, might still outperform the SB-approach, however do not allow for separate updates of the system's variables and are therefore less suited for parallel computing. 

The SB-algorithm is, to our knowledge, currently the fastest way to solve a fully connected Ising problem and therefore also an ideal candidate to solve the optimal integer portfolio problem in the Ising representation.

\section{Trading Trajectory Problem as an Ising Problem}
In this section the optimal trading trajectory problem is formulated and mapped to a binary-bit representation and then mapped to the Ising representation.
\subsection{Integer Representation}
The optimal trading trajectory problem can be described as a temporal sequence of mean variance portfolios \cite{markowitz1, markowitz2, markowitz3}, which are composed of an expected return, an expected risk and an expected trading/rebalancing cost term. Mathematically this can be expressed by
\begin{equation}
\label{eq:optimalporfolio}
w = \argmax_w \sum_t \left( w^T_t \mu_t - \frac{\gamma}{2}w^T_t \Sigma_t w_t  -\Delta w_t^T \Lambda_t \Delta w_t \right), 
\end{equation}
where $w_t$ is the portfolio weight vector of size $N$ at time $t$ and the individual entries fulfil $w_{it} \in \mathbb{N}_0$. The vector $\mu_t$ is the estimated future return vector for time $t$, $\Sigma_t$ the estimated covariance and $\gamma$ a risk aversion parameter. The matrix $\Lambda_t$ holds the costs to perform the rebalancing of investments from time point $t$ to $t+1$. 

Often the optimization is formulated with a set of constraints. A common constraint that we will include is that the weights are subject to $\sum_{i=1}^{N} w_{it} \leq \tilde{K}$ at all time-points, where $\tilde{K}\geq 0$ is the total available units to be spent. For simplicity we assume that $\tilde{K}$ is constant over time. We will introduce a further constraint that allows only for limited investments in each asset. This constraint could also be time dependent and different for each asset, for simplicity we however choose a time independent and uniform constraint of the form $\forall t: \, \, 0 \leq w_{it} \leq \tilde{K}/N$, that allows only a certain fraction of our total units (total capital) to be invested in one asset.

\subsection{Map Optimal Portfolio Problem to Ising Model}
Both, the Ising problem and the portfolio problem, at least in their native form, have the same structure and ask us to minimize an expression with a linear and a quadratic term. The formulation with spins, where $s_i \in \{-1,1\}$, is closely related to the perhaps more intuitive and well known binary bit representation where $b_i \in \{0, 1\}$, which can be obtained by $b_i = (s_i+1)/2$. Therefore we first map the portfolio problem with integer weights to the binary-bit representation and then to the binary-spin representation. Mapping the optimal porfolio problem to a binary representation has already been conducted multiple times using various binary representations \cite{dwaveportfolio}. 

Let us first move to the binary representation by rewriting $w_{it} = \sum_{k=0}^{\log_2(K)+1} 2^{k}b_{ikt}$ where $b_{ikt} \in \{0, 1\}$ in which $K$ is the largest integer we want to represent in $w$ (note the difference to $\tilde{K}$). All individual bits $b_{ikt}$ can be packed into a single vector $b$, with which eq. \ref{eq:optimalporfolio}, dropping $\Lambda_t$ for now, can be reformulated as
\begin{equation}
\label{eq:optimalporfolio_binary}
b = \argmax_b \sum_t \left( b^T_t \hat{\mu}_t - \frac{\gamma}{2}b^T_t \hat{\Sigma}_t b_t \right), 
\end{equation}
where the hat simply indicates that these are the respective quantities in the new basis. The future return vector and the covariance matrix will need to be extended and appropriate entries multiplied with $2^d$, the exact order and value depends on how the elements of $b$ are ordered. Further the length of $b$ is larger than $w$ by a factor of $\floor{\log_2(K)}+1$. In order to eliminate all ambiguity let us for clarity write out all the sums that are hidden in the vector multiplication above.
\begin{equation}
b = \argmax_b \sum_t \left( \sum_{i} \sum_{k} 2^k b_{ikt} \mu_{it} - \frac{\gamma}{2} \sum_{i,j} \sum_{k,l} 2^k 2^l b_{ikt}b_{ilt}\Sigma_{ijt}  \right), 
\end{equation}
where the sum over $k$ and $l$ starts at 0 and goes until $\floor{\log_2(K)}+1$ and $i$ and $j$ are the indices labelling all assets. The term $\Sigma_{ijt}$ stands for the covariance between asset $i$ and $j$ at time $t$. We chose to order $b$ according to the bit endianness, meaning bundling by significance of respective bit. In this basis the transformed future returns $\hat{\mu}$ and $\hat{\Sigma}$ can be easily constructed by a trivial block-wise extension of the original quantities found in eq. \ref{eq:optimalporfolio}. Each block (and respective entries) $\hat{\Sigma}_{klt}$ are simply obtained by multiplying the $\Sigma_t$ matrix with the corresponding bit significance, i.e. $2^k 2^l \Sigma_t$, which is an $N\times N$ matrix block. For the linear term $\hat{\mu}_t$ we simply expand $\mu_t$ and multiply the respective entries with $2^k$. We have picked one binary representation and also one distinct choice of basis, note however that there are many other options with their advantages and disadvantages. An interesting overview on different binary representations was part of the study \cite{dwaveportfolio} around the portfolio optimization on a D-Wave system.

The portfolio problem in its binary-bit basis in eq. \ref{eq:optimalporfolio_binary} can now be transform to the spin representation by introducing $b=(s+\mathbf{1})/2$, where $\mathbf{1}$ denotes the unit vector with same length as $s$. This introduces scalar correction terms independent of $s$, for the optimization step however they can be dropped and the overall expression can be compressed to
\begin{equation}
s = \argmin_s \sum_t \left[ \frac{\gamma}{4} s^T_t\hat{\Sigma}_t s_t  + \left(\frac{\gamma}{2}\mathbf{1}\hat{\Sigma}_{t} - \hat{\mu}\right)s_t \right] \quad ,
\end{equation}
where we also flipped the sign in order to minimize the expression. From here we can directly read off the corresponding couplings of the respective problem in the Ising representation. The quadratic blocks (in our basis) appearing in the block diagonal matrix $J$ are $J_t = -\frac{\gamma}{2}\hat{\Sigma}_t$. The linear term in the Ising representation can also be broken down into temporal sequences that are obtained via $h_t = \frac{\gamma}{2}\mathbf{1} \hat{\Sigma}_{t}- \hat{\mu_t}$. Note that the block diagonal nature of $J$ is extended with further blocks linking spins of different time points with each other, as soon as the transaction costs in $\Lambda_t$ are introduced. These blocks in the Ising representation $\hat{\Lambda}_t$, linking spins from different time-points can be constructed such that they introduce a tendency for identical spins on neighbouring time points to be aligned, unless the contribution of the expected future return is strong enough to trigger a spin flip. A way how this can be achieved is to add terms of the form $\frac{1}{2}s_{ik,t}s_{ik,t+1}c_i(t,t+1)2^{k}$ to the sum, where $i$ denotes the asset, $k$ the bit-significance of the respective spin and $c_i(t,t+1)$ the transaction penalty. We will follow exactly this approach, will however choose  a constant cost $c_i(t,t+1)=c$ for all assets and time points. Like this the diagonal matrices that appear in off-diagonal blocks of the interaction matrix J, reduce to $\hat{\Lambda}_t=\frac{1}{2}\mathbf{1}c$, where we omitted writing the multiplication of the spin dependent factors of 2. The just discussed approach is one way to generate energy-gaps in units of $c$ depending only on $\Delta w$.

To include constraints in the Ising representation can be more elaborate than in the integer representation of the problem. Various different constraints can for example demand the introduction of ancillary spins \cite{ancillaspins}, this is why we chose a set of constraints that are directly embedded as hard constraints in the formulation of the problem. If for a given asset $i$ we choose to have only $\floor{\log_2(\tilde{K}/N)}+1$ different spins, we automatically embed the constraint $w_{it} \leq \tilde{K}/N$ and hence also $\sum_{i=1}^{N} w_{it} \leq \tilde{K}$.

\section{Simulated Bifurcation of Ising Problem}
In this section we discuss the core components of the simulated bifurcation algorithm, introduced in a recent study \cite{simulatedbifurcation}, from which we will borrow heavily in this section (content and also nomenclature) . 

\subsection{Classical Description of Quantum Adiabatic Evolution}
Following the exact steps from \cite{simulatedbifurcation} we formulate the Ising energy defined in eq. \ref{eq:isingenergy} with a network of Kerr non-linear parametric oscillators. In a quantum mechanical formulation the Hamiltonian is given by 
\begin{eqnarray}
\label{eq:hamiltonianquantum}
\hat{H}(t) = &\hbar & \sum_i \left( \frac{K}{2}a_i^\dagger a_i^2 - \frac{p(t)}{2}(a_i^{\dagger 2} + a_i^{2}) + \Delta_i a_i^\dagger a_i \right)\nonumber \\
&-&\hbar \xi_0 \sum_{i,j} J_{ij} a^\dagger_i a_j + \hbar\xi_0 A(t) \sum_i h_i (a^\dagger_i + a_i) \, ,
\end{eqnarray}
where $a^\dagger_i$ is the creation and $a_i$ the annihilation operator for the i-th oscillator. The parameter $\Delta_i$ is the detuning frequency which plays an important role when defining the initial vacuum state. The time dependent parameter $p(t)$ is the pumping amplitude, $\xi_0$ is a constant parameter (in units of frequency) and $K$ is the Kerr coefficient. Due to the same arguments given in \cite{simulatedbifurcation} we assume that $K, \Delta$ and $\xi_0$ are positive. For our considerations the numerical value of the reduced Planck constant $\hbar$ is irrelevant, note however that it also carries the unit of time which gets cancelled with the unit of frequency (inverse time) in $\xi_0$, guaranteeing that the entire expression is in units of energy only. $A(t)$ is a positive dimensionless parameter that increases with $p$ over time, such that $A \approx 0$ when $p \ll \Delta$ and $A \approx \sqrt{(p-\Delta)/K}$ when $p \gg \Delta$.

A quantum adiabatic evolution of this Hamiltonian is desired which will end up in the ground-state that will minimize the Ising energy. In order to achieve this we initialize all oscillators in their vacuum states and gradually increase the pumping amplitude $p(t)$ from zero to a sufficiently large value compared to $\Delta$ and $\xi_0$. In order to initialize the system in the vacuum state a proper tuning of  $\Delta_i$ can be necessary. The reader is referred to H. Goto's work \cite{vaccumsateGoto} for more details on the initialization of the vacuum-state. Also finding a proof in the appendix that shows that if the variation of $p(t)$ is sufficiently slow, the final state will become the ground-state of the final Hamiltonian by the quantum adiabatic theorem.

In the following we will formulate the corresponding classical Hamiltonian which can be derived by approximating the expectation values of $a_i$ via a complex amplitude $x_i + iy_i$ (note that introductory literature often uses $p$ and $q$ here instead). The real and imaginary part form a conjugate variable pair that correspond to position and momentum of the i-th oscillator. With $x_i = (a_i+a_i^\dagger)/2$ and $y_i =(a_i - a_i^\dagger)/2 $ we can describe eq. \ref{eq:hamiltonianquantum} with a classical expression of the form
\begin{eqnarray}
\label{eq:classicHamiltonian}
H(x,y,t) &=& \sum_i\left[ \frac{K}{4}(x^2_i+y^2_i)^2 - \frac{p(t)}{2}(x^2_i -y^2_i) + \frac{\Delta_i}{2}(x^2_i+y^2_i) \right] \nonumber \\
 &-& \frac{\xi_0}{2}\sum_{ij}J_{ij}(x_ix_j + y_iy_j) + 2\xi_0 A(t)\sum_i h_i x_i 
\end{eqnarray}
Note that in this formulation, after adiabatic evolution, the spin value $s_i$ corresponds to the sign of the amplitude $x_i$. Since eq. \ref{eq:classicHamiltonian} describes a classical Hamiltonian we can derive the equations of motions for variables by taking the derivative with respect to time, denoted with a dot, these are given by following the classical time evolution formulas of Hamiltonian mechanics:
\begin{eqnarray}
\label{eq:motionX}
\dot{x}_i = \frac{\partial H}{\partial y_i} =& & \left[K(x^2_i +y^2_i) + p(t) + \Delta_i \right]y_i \nonumber \\
& -& \frac{\xi_0}{2}\sum_j J_{ij}y_j\\
\label{eq:motionY}
\dot{y}_i = -\frac{\partial H}{\partial x_i} = &-& \left[K(x^2_i +y^2_i) - p(t) + \Delta_i \right]x_i \nonumber \\
&+& \xi_0 \sum_j J_{ij} x_j - 2\xi_0A(t)h_i
\end{eqnarray}
Note that there is a minus sign in front of the derivative of the second conjugate variable. These formulas describe kinetics that allow us to simulate classically the quantum adiabatic evolution.

\subsection{Simulated Bifurcation Algorithm}
The equations of motion, derived in eq. \ref{eq:motionX} and \ref{eq:motionY}, can be further simplified in order to be more suitable for fast numerical simulation. Again following \cite{simulatedbifurcation} the terms proportional to the momenta $y$, which vary around zero, can be dropped and the equations of motion can be reformulated as
\begin{eqnarray}
\label{eq:motionXapprox}
\dot{x}_i &= &  \Delta_i y_i \\
\label{eq:motionYapprox}
\dot{y}_i &=& - \left[K x^2_i - p(t) + \Delta_i \right]x_i \nonumber \\
& &+ \xi_0 \sum_j J_{ij} x_j - 2\xi_0A(t)h_i
\end{eqnarray}
Note that this approximation allows us to use the symplectic Euler method \cite{symplectic} to simulate the hamiltonian dynamics of the system because the two variables are now separable. We therefore discretize time with $t=n\Delta_t$, where $\Delta_t$ is our time increment. Like this we can write the algorithmic update step for the position variable as 
\begin{equation}
x_i(t_{n+1}) = x_i(t_n) + \Delta_i y_i(t_n) \Delta_t
\label{eq:sbalgo0}
\end{equation}
and the update of the momentum variable as
\begin{align}
  y_i(t_{n+1}) &= y_i(t_n) - \left[ K x^3_i(t_{n+1}) + (\Delta_i -p(t_{n+1}))x_i(t_{n+1}) \right. \nonumber \\
  &\qquad - \left. \xi_0 \sum_j J_{ij} x_j(t_{n+1}) + 2 \xi_0 A(t_{n+1}) h_i \right] \Delta_t \, .\label{eq:sbalgo}
\end{align}
From here a number of further different simplifications and re-formulations can be made in order to arrive at an even faster algorithm. The expression above however describes the core update steps of the SB algorithm.

\section{Results}
The results section is divided into four smaller sub-sections. First we will discuss results obtained for portfolios at a single time point, then present solutions for trading trajectory problems, then discuss performance and finally consider problems where only close-to-optimal solutions were found.

\subsection{Optimal Portfolio with SB-Algorithm}
In this section we consider results, obtained by the SB-algorithm, for the optimal portfolio problem formulated in the Ising representation. For the results in this section we create an artificial market situation with $N$ different assets by sampling random returns from a geometric Brownian motion over 1000 time increments in order to estimate a future return vector $\mu$ and a covariance matrix $\Sigma$. Optionally we might include a drift in the market that results in an average expected return of $\kappa$. 

\begin{figure}[h]
 \includegraphics[width=0.5\textwidth, angle=0]{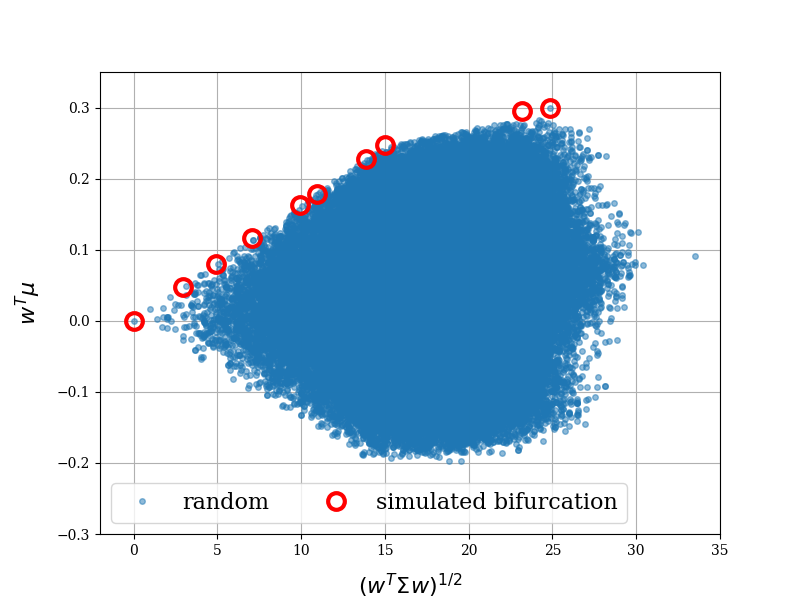}
 \caption{Shows $10^5$ random portfolios (including the important edge cases) and those found via adiabatic evolution in the simulated bifurcation algorithm for various different risk aversion parameters $\gamma$. This is for an artificial portfolio with $N=5$ assets, a total capital of $\tilde{K}=75$ units, where $15$ can be spent maximally per asset. The average expected return of the sampled returns is 0.5\%}
\label{fig:small}
\end{figure}

In fig. \ref{fig:small} we show the configurations found by the SB-algorithm for a small portfolio of $N=5$ different assets, a total amount of $\tilde{K}=75$ units available and a maximal amount of $\tilde{K}/N=15$ units to be distributed per asset. Further $10^5$ random portfolios are added, with the same constraints, to illustrate the universe of different investment options. This is done in a market with a positive drift resulting in an average expected return of $\kappa = 0.5\%$. The results in fig. \ref{fig:small} show that the SB-algorithm optimized the portfolio allocation correctly, i.e. for a given risk picks the configuration which maximizes the return. The various points were constructed by increasing $\gamma$ from zero to a sufficiently large value.

\begin{figure}[h]
 \includegraphics[width=0.5\textwidth, angle=0]{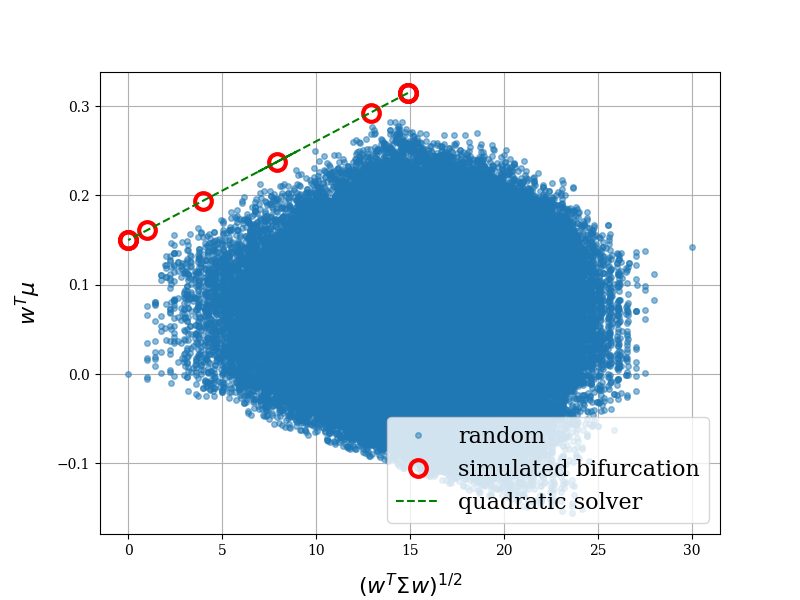}
 \caption{Shows $10^5$ random portfolios (including some edge cases) and those found via adiabatic evolution in the simulated bifurcation algorithm for various different risk aversion parameters $\gamma$. This is for an artificial portfolio with $N=5$ assets, a total capital of $\tilde{K}=75$ units, where $15$ can be spent maximally per asset. The average value of the sampled returns is 0\%, but we introduced a risk free asset with 1\% return, visible by the off-set on the y-axis. Further the optimal portfolio boundary is highlighted in green.}
\label{fig:small_quadraticsolv}
\end{figure}

\begin{figure}[h]
 \includegraphics[width=0.5\textwidth, angle=0]{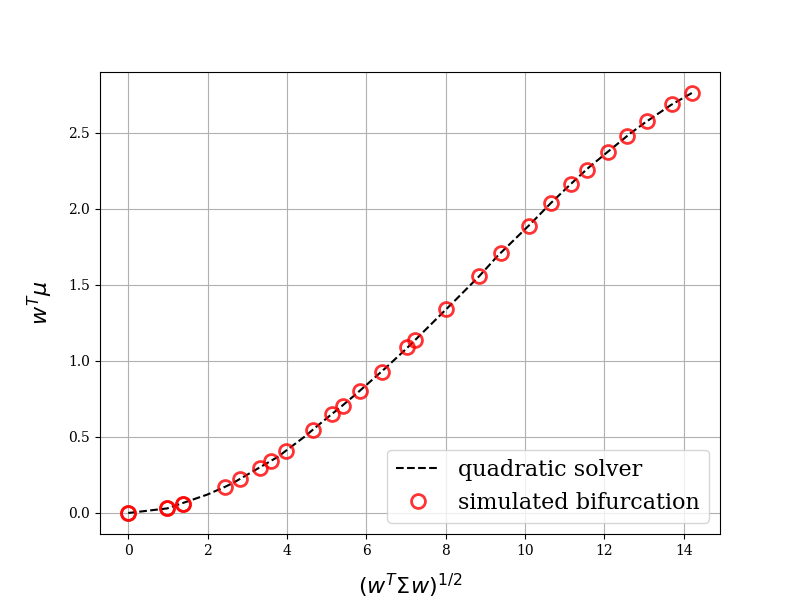}
 \caption{Portfolio with $N=400$ assets, a total capital of $\tilde{K}=400$ units, where $1$ can be spent maximally per asset, either purchase or not. The average expected return of the sampled returns is 0.5\%. Computation time of the approximation via the quadratic solver of the continous problem and the SB-approach were identical, both around 0.3 seconds per point on a single Intel Core i5-7200.}
\label{fig:verylarge_quadraticsolv}
\end{figure}

We repeat this experiment and replace a random asset with a risk free asset with 1\% future return. If we push the risk aversion parameter $\gamma$ to large enough values we should end up in the scenario where we invest only in the risk free asset and avoid any investments in risky assets. Exactly this result is shown in fig. \ref{fig:small_quadraticsolv}, where we see that the SB-algorithm correctly finds the optimal portfolios (as also highlighted by the optimal frontier line calculated via a quadratic solver), and ends up in the just described risk free scenario when $\gamma$ reaches large enough values. Since $\tilde{K}/N=15$ and the risk free asset's return is 1\%, the risk free portfolio corresponds to the point (0, 0.15), as shown in the figure.

In fig. \ref{fig:verylarge_quadraticsolv} we are considering a scenario where the number of assets is $N=400$ and we can either buy one unit of the asset or not. This allows for very fast computation times of less than half a second per point on a standard desktop CPU. For the extreme case of $N=1000$ and $\tilde{K}/N=1$, the SB-algorithm uses roughly $1$ seconds to find the optimal solution, which even beats many out of the box quadratic solvers to tackle the respective continuous problem with the identical constraints. We investigate the performance in more detail in section \ref{sec:performance} and in the next section consider the solutions found by the SB-approach for the optimal trading trajectory problem.

\subsection{Optimal Trading Trajectory with SB-algorithm}
\label{sec:opttradingtraj}
We now construct $T$ future random market conditions with different values of average expected returns and use the SB-algorithm to find the optimal asset allocation trajectory over time. We first consider a setup with only $N=3$ and $\tilde{K}/N=3$, but $T=100$, which already generates $2^{600}$ different trajectories from which we want to find the optimal one. The trading costs are organised such that for a fixed asset we get a cost penalty of $c=0.01$ for each unit changed between two time steps. We will consider the optimization for three different $\gamma$ values and consider the trajectories found by the SB-algorithm. 

\begin{figure}[h]
 \includegraphics[width=0.5\textwidth, angle=0]{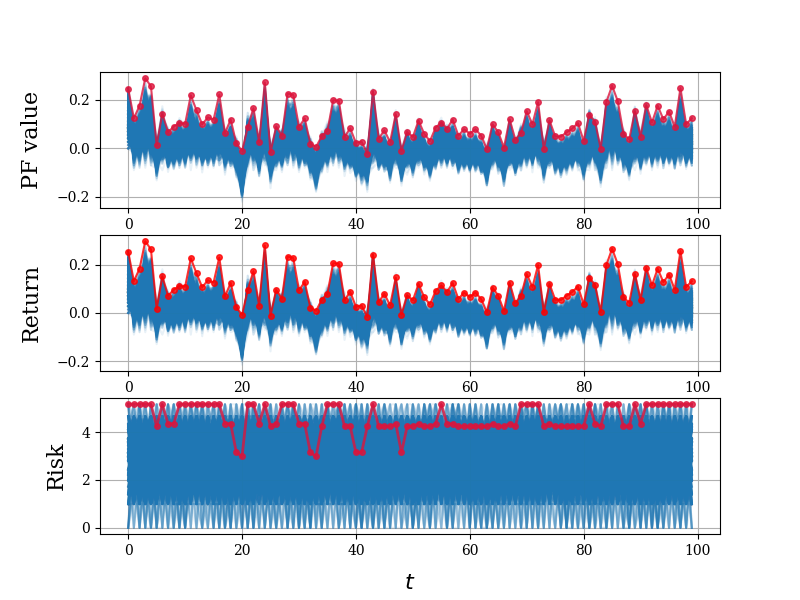}
 \caption{Optimal trajectory for $N=3$, $\tilde{K}/N=3$ and $T=100$ for zero and close to zero risk aversion.}
 \label{fig:trajectory_smallGamma}
\end{figure}

In fig. \ref{fig:trajectory_smallGamma} at the top we show the trajectory value $w_t^T\mu_t-\frac{\gamma}{2} w^T_t \Sigma_t w_t -\Delta w_t^T \Lambda_t \Delta w_t$ at each time point $t$. The red line again indicates the trajectory found by the SB-algorithm, whereas the blue samples indicate the universe of possible trajectories. In the second row of fig. \ref{fig:trajectory_smallGamma} only the return term $w_t^T\mu_t$ is shown and in the bottom only the risk component $w^T_t \Sigma_t w_t$. For $\gamma=0$, and also for sufficiently small values of the risk aversion parameters, the optimization procedure can ignore the risk, which is exactly what is observed when looking at the solution found by the SB-algorithm. As we see in the bottom of fig. \ref{fig:trajectory_smallGamma}, the trajectory is exposed to a lot of risk and solely focuses on optimizing the return and trading costs.

\begin{figure}[h]
  \includegraphics[width=0.5\textwidth, angle=0]{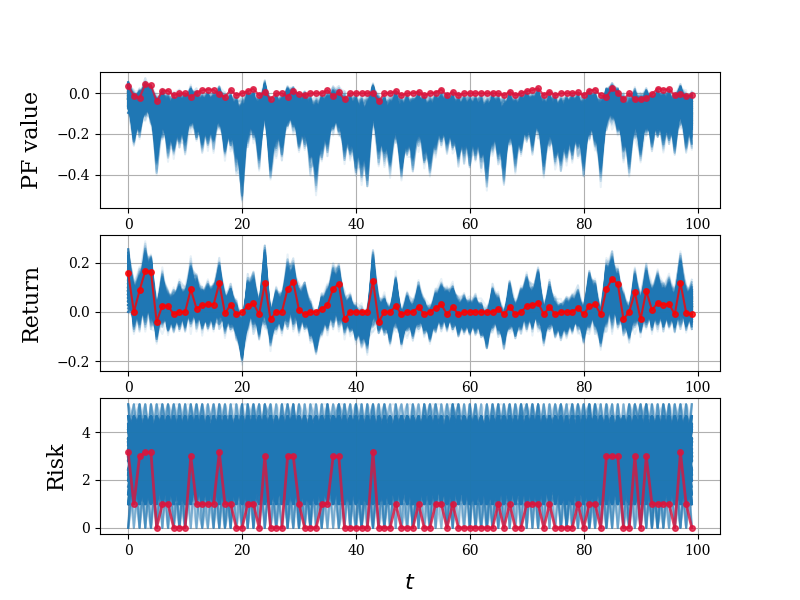}
 \caption{Optimal trajectory for $N=3$, $\tilde{K}/N=3$ and $T=100$ for $\gamma = 0.02$.}
\label{fig:trajectory_mediumGamma}
\end{figure}

In fig. \ref{fig:trajectory_mediumGamma}, we are looking at the same scenario as in fig. \ref{fig:trajectory_smallGamma}, this time however, with an increased risk aversion parameter set to $\gamma=0.02$. This change makes us more risk averse and we observe how the SB-algorithm finds a trajectory  for which risk is only taken if the magnitude of expected return is large enough, else, if the risk term dominates, portfolios are found that have a sufficiently small risk value. In the top of fig. \ref{fig:trajectory_mediumGamma}, we see that the overall portfolio value, including return and properly weighted risk still looks meaningfully maximized. 

\begin{figure}[h]
  \includegraphics[width=0.5\textwidth, angle=0]{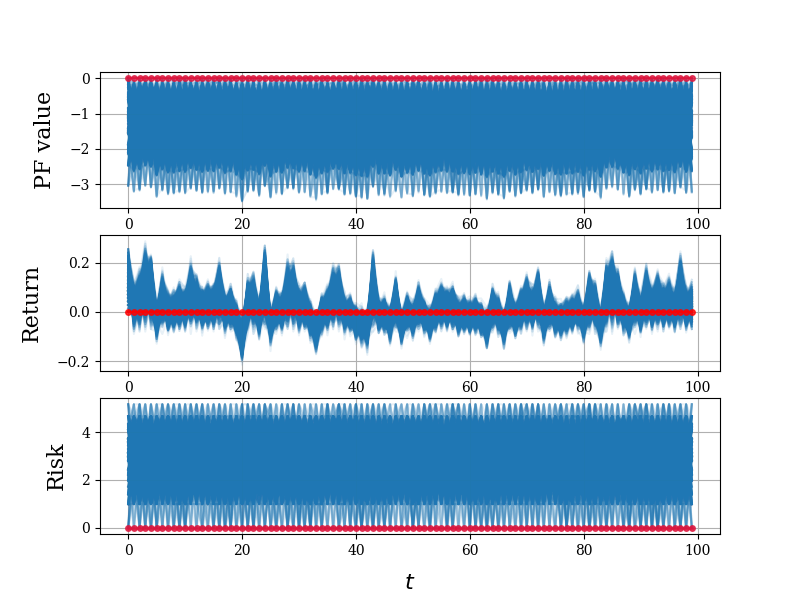}
 \caption{Optimal trajectory for $N=3$, $\tilde{K}/N=3$ and $T=100$ sufficiently large risk aversion parameters.}
 \label{fig:trajectory_largeGamma}
\end{figure}

In fig. \ref{fig:trajectory_largeGamma} the risk aversion parameter was increased further in order to check if we observe the extreme scenario in which the risk term is dominating completely and forcing a zero investment trajectory. This is exactly what we observe when looking at the solution found by the SB-algorithm, risk is minimized completely by not suggesting any investments during the entire time period. In the middle panel of fig. \ref{fig:trajectory_largeGamma} we again see the potential future return, from which we however not benefit due to the increased risk awareness.

\begin{figure}[h]
  \includegraphics[width=0.5\textwidth, angle=0]{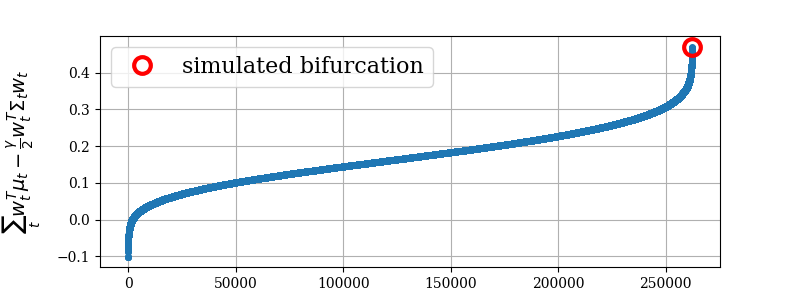}
 \caption{Shows sorted values of all possible $2^{18}$ trajectories in a setup where $N=3$, $\tilde{K}/N=3$ and $T=3$. The In addition the configuration obtained with the SB-algorithm is plotted, giving numerical evidence of finding the global optimum for the mentioned system. Note that the corresponding trading costs of each trajectory have been included in the total portfolio value.}
 \label{fig:smallEvidence}
\end{figure}

The SB-algorithm finds trajectories under different conditions that are meaningful and that seem to coincide with the expected results, we have however not delivered proof that, at finite trading costs, the found solutions correspond to the true optimal trajectory. Let us in the following establish numerical evidence that the SB-algorithm finds the trajectory of configurations that correspond to the global optimum. For this we consider a smaller system with only 3 time points. In this small system it is possible to compare all possible trajectories with the solution obtained by the SB-algorithm. In fig. \ref{fig:smallEvidence} we show the ordered accumulated portfolio values of all $2^{18}$ possible trajectories for a system with $N=3$, $\tilde{K}/N=3$, $T=3$ and $c=0.01$. We observe that the SB-algorithm finds the true global optimum out of all configurations.

\begin{figure}[h]
  \includegraphics[width=0.5\textwidth, angle=0]{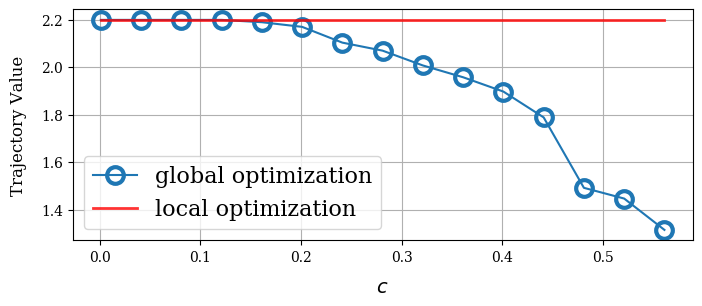}
 \caption{The blue line shows the total trajectory value of the solutions of the global optimization problem for different finite trading costs. The red line indicates the value obtained when optimizing every portfolio at each point in time separately, i.e. performing local optimization. This is for a system with $N=3$, $\tilde{K}/N=3$ and $T=100$.}
\label{fig:trajectory_numericalcheck}
\end{figure}

If we increase the temporal dimension to $T=100$, we already have $2^{600}$ possible combinations, where our comparison with a brute force method is infeasible and the comparison to random trajectories becomes meaningless. For those larger systems we were only able to check the correct behaviour of the approach in certain limits. In the limit of vanishing trading costs for example, i.e. $c \to 0$, the system reduces to a time independent optimization problem of $T$ separately solvable portfolios. The numerical checks in that limit show that the global optimization coincides with the trajectory that optimizes the portfolios at each point in time separately. This is illustrated in fig. \ref{fig:trajectory_numericalcheck}, where we see how the globally optimized portfolio converges to the local optimization when decreasing the trading penalty. 

\begin{figure}[h]
  \includegraphics[width=0.5\textwidth, angle=0]{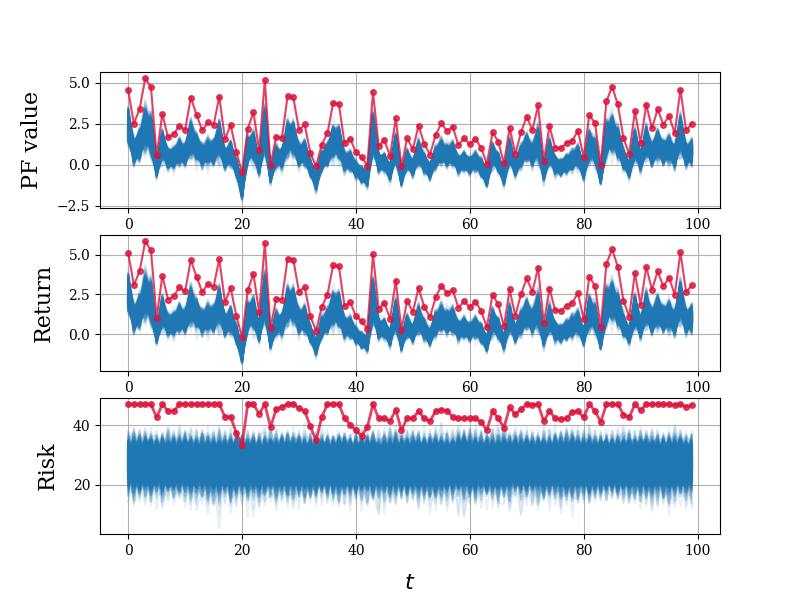}
 \caption{Optimal trajectory for $N=10$, $\tilde{K}/N=15$ and $T=100$.}
\label{fig:trajectory_largeSystem}
\end{figure}

Numerical checks are only possible in certain limits, and hence we can not guarantee the correctness of solutions found for much larger random markets with non-zero trading costs. The confirmation obtained by comparing smaller systems with brute force methods and the correct behaviour of larger systems in given limits, give us confidence, however not proof, that we find optimal or close-to-optimal results for larger systems with finite trading costs. Another example where a direct cross check with another method was not possible is illustrated in fig. \ref{fig:trajectory_largeSystem}, where we optimize a portfolio with $N=10$, $\tilde{K}/N=15$ and $T=100$, which generates $2^{4000}$ possible combinations. 

Note that for some scenarios we observed certain exceptions where the SB-algorithm only found close-to-optimal solutions, those will be discussed in section \ref{sec:problems}.

\begin{figure}[h]
 \includegraphics[width=0.6\textwidth, angle=0]{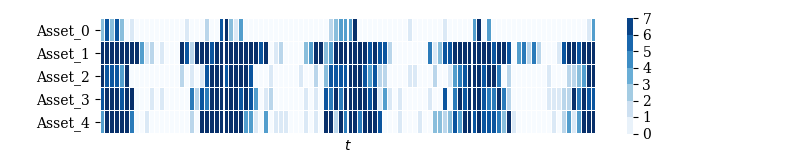}
 \includegraphics[width=0.6\textwidth, angle=0]{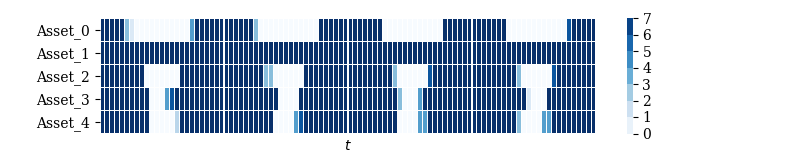}
 \caption{Trading trajectory investment suggestion from the SB-algorithm for $N=4$, $\tilde{K}/N=7$ and $T=100$ with a cross-asset seasonal and a random component in the expected future market. On top of the random component a seasonal affect is included. In the top panel the trading cost term is set to $c=0.001$ allowing for frequent trading, whereas in the bottom to $c=0.04$, making any change extremely costly.}
 \label{fig:investmentsuggestion}
\end{figure}

In order to gain insight into the optimal trading strategy of a selected trajectory we can display the number of suggested units to be held for each asset for each point in time. A simple illustration of this with a portfolio of $N=4$, $\tilde{K}/N=7$ and 100 different time points is shown in fig. \ref{fig:investmentsuggestion}. From the top panel to the bottom panel we increase the trading penalty term $c$ and, as expected, observe that the optimal trajectory found by the SB-algorithm avoids changing the positions as frequently as before. The future market was generated with a seasonality and a random effect in order to enforce also seasonal effects in the trading pattern. The result in fig. \ref{fig:investmentsuggestion} establishes confidence that the penalty term introduced in the Ising representation allows to control the trading costs, e.g. the amount of trades. This however is only a heuristic sanity check and numerical evidence which proves that the SB-algorithm found the global optimum including trading costs were only possible in the already discussed setups. Despite the successfully controllable trading activities, the translation from real world trading costs and concepts to properly calibrated penalty terms in the Ising formulation still forms an open challenge that is subject of ongoing investigations.

\subsection{Performance}
\label{sec:performance}
\begin{figure}[h]
 \includegraphics[width=0.5\textwidth, angle=0]{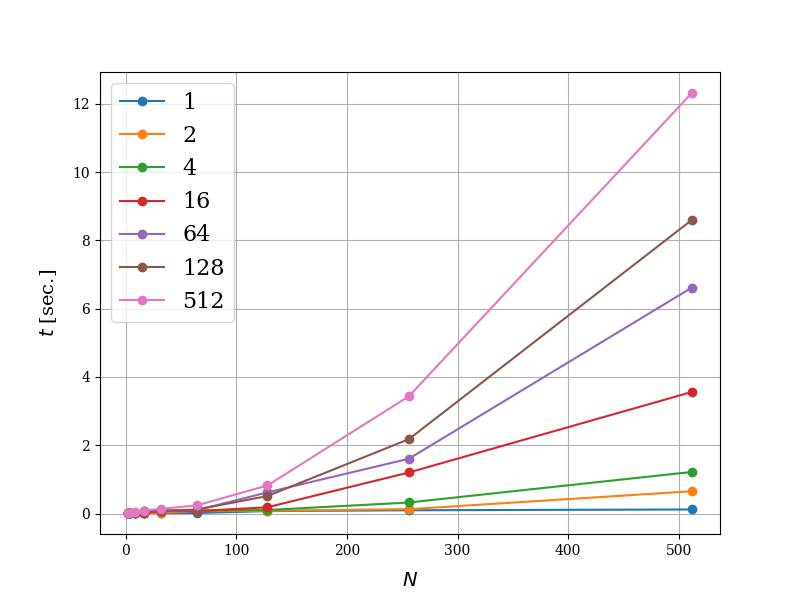}
 \caption{Shows the computation time of the SB-algorithm for increasing number of assets $N$. The different lines correspond to different values of maximal investments per asset. Each point is estimated by averaging 10 identical experiments to account for eventual fluctuations. This was done for a fixed set of SB-parameters $\epsilon_p=0.01, \Delta_t=0.01, \Delta_i=1$.}
 \label{fig:performance}
\end{figure}

Goto et. al. \cite{simulatedbifurcation} took advantage of the fact that the SB-algorithm can be set up in a highly parallel manner on a GPU cluster and thereby solved a fully connected 2000 spin problem $10$ $\times$ faster than the current state of the art custom built laser machine. We are following the same path for the portfolio optimization by porting the discussed SB optimization procedure to a GPU cluster. Performance numbers based on the GPU implementation are currently not available, we will however consider the already convincing performance on a single Intel Core i5-7200 with 2.5GHz (fully vectorized eq. \ref{eq:sbalgo0} and eq. \ref{eq:sbalgo} in Python 3.7). 
For this we measure the time of the adiabatic evolution in relation to the system size. For comparison we keep the parameters of the SB-algorithm ($p_{\text{max}}, \epsilon_p, \Delta_t, \Delta_i$) fixed, when increasing the number of assets $N$ and the number of maximal units per asset $\tilde{K}/N$. In fig. \ref{fig:performance} we see the level of increase in computation time when increasing $N$. Despite the fact that the computation time shows an exponential increase with increasing system size, it is astonishing to observe that the simulated bifurcation of a system with 256 assets, with maximal investment of 512 units per asset is performed under 4 seconds. It cannot be directly compared with other methods that ran in different setups (e.g. different constraints), it is however noteworthy to mention that it clearly beats the currently existing competitive numbers, like the 200 asset optimization by a branch and bound method that showed an average run-time of 4800 seconds \cite{exactintegersolution}. The move to a GPU cluster will not only allow to drastically decrease the already fast computation time but will also allow to include an extremely large amount of assets.

\begin{figure}[h]
 \includegraphics[width=0.5\textwidth, angle=0]{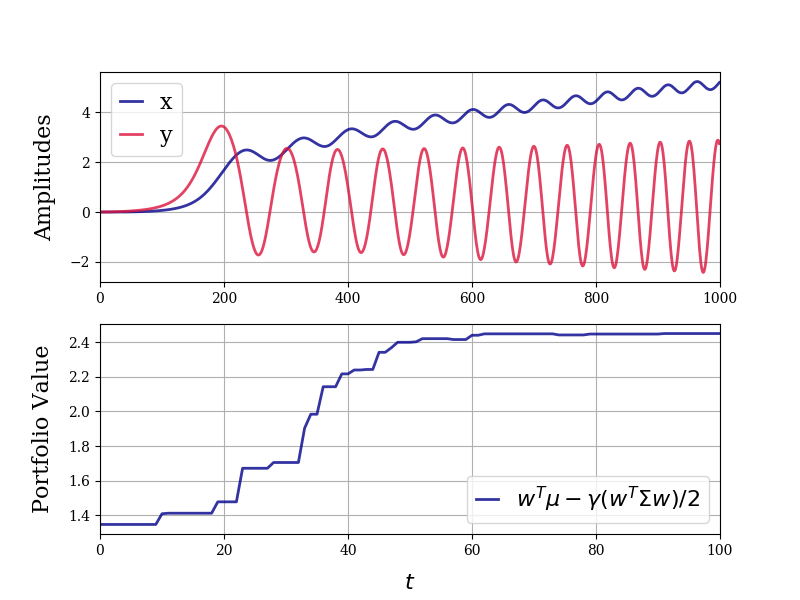}
 \caption{Top: Shows the average of the position amplitude $x$ and momentum amplitude $y$ over the time in which the pumping amplitude $p(t)$ is increased. Bottom: Shows the portfolio value during the first part of $t$. This is for $N=20$, $\tilde{K}=300$, $\gamma = 0.001$, $\epsilon_p=0.1$ and $\Delta_t = 0.01$.}
 \label{fig:evolution}
\end{figure}

The measurements displayed in fig. \ref{fig:performance} create a valuable insight into the proper scaling of the algorithm, it however is not guaranteed that a ground-state is found in the displayed time for all systems of that size. For smaller systems the parameters in the algorithm could be set much more aggressive, resulting in an even faster simulated bifurcation that would still result in the proper ground-state. In fig. \ref{fig:evolution} we show the amplitudes of the conjugate variables describing position and momentum of the approximated Kerr-oscillators in the top panel and the maximized portfolio value in the bottom panel during the time in which the pumping amplitude $p(t)$ is increased. In this setup, where the number of assets are only 20, the optimal value has been reached before the pumping amplitude has reached its maximal value, theoretically allowing us to reduce computational efforts whilst still finding the same state. Note that this effect also goes into the other direction and for larger systems the properly tuned parameters can lead to an increase in computation time.

For large enough systems, or scenarios with extreme values of $\mu$ and $\Sigma$, a fixed set of parameters will eventually lead to meaningless results. As expected we observed that in scenarios where the steps of the pumping amplitude $\epsilon_p$ and the finite time increments $\Delta_t$ are picked too large, the algorithm will not end up in the ground-state. In such a scenario the 'resolution' of the algorithm needs to be increased and hence the respective computation time to simulate the adiabatic evolution will increase as well. The exact parameter choice has proven to be a delicate fine-tuning problem which, if done properly however, can also help reduce computation time. 

Note that we have presented the performance measurements from an end-user perspective and will skip a detailed discussion around the bifurcation phenomenon and the probability of success of the algorithm here. A highly valuable discussion around such can be found in the original introduction \cite{simulatedbifurcation} of the algorithm.

\subsection{Close-to-Optimal Solutions}
\label{sec:problems}

Certain systems can have multiple configurations with Ising energy values, or equivalently portfolio values, that are numerically very close to each other or even identical. The level of degenerate and almost degenerate states depends on the number of assets $N$, the amount of units to distribute $\tilde{K}$ and on the expected return and risk of the assets. The amount of degenerate and almost degenerate configurations increases for larger values of $\tilde{K}$, particularly if the estimated returns and risks are very similar among the assets. What we observe is that for systems with larger values of $\tilde{K}$, the SB-algorithm needs a much more carefully picked set of parameters to find the optimum. If the parameters of the SB-algorithm are not tuned to an optimum, the algorithm seems to fail to detect those seemingly in-existing energy differences among the different configurations and ends up in one of the many (pseudo-)degenerate ground-states, depending on the initial random configurations of the oscillators. In the following we will discuss an example where, without delicate fine-tuning of the SB-parameters, the SB-algorithm finds only close-to-optimal solutions, and where the increased level of degeneracy is expected to be strongly connected to the issue. Note however, that even for systems with a high level of degeneracy, we can still find close-to-optimal solutions for very large $N$ and $\tilde{K}$ faster than approaches from previous studies, as was illustrated in fig. \ref{fig:performance}.

\begin{figure}[h]
  \includegraphics[width=0.5\textwidth, angle=0]{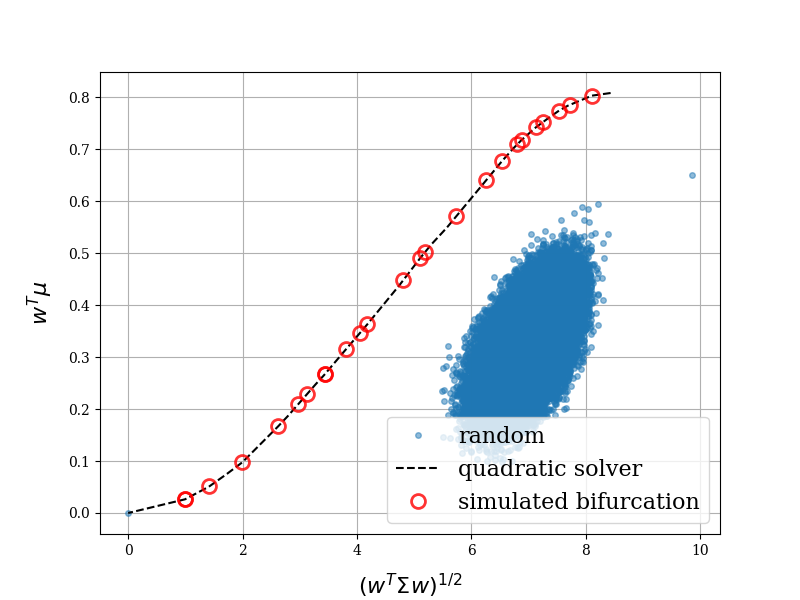}
 \caption{$N=100$, $\tilde{K}/N=1$ for various values of $\gamma$. Additionally $10^6$ random portfolios are shown.}
\label{fig:N100K1}
\end{figure}

\begin{figure}[h]
  \includegraphics[width=0.5\textwidth, angle=0]{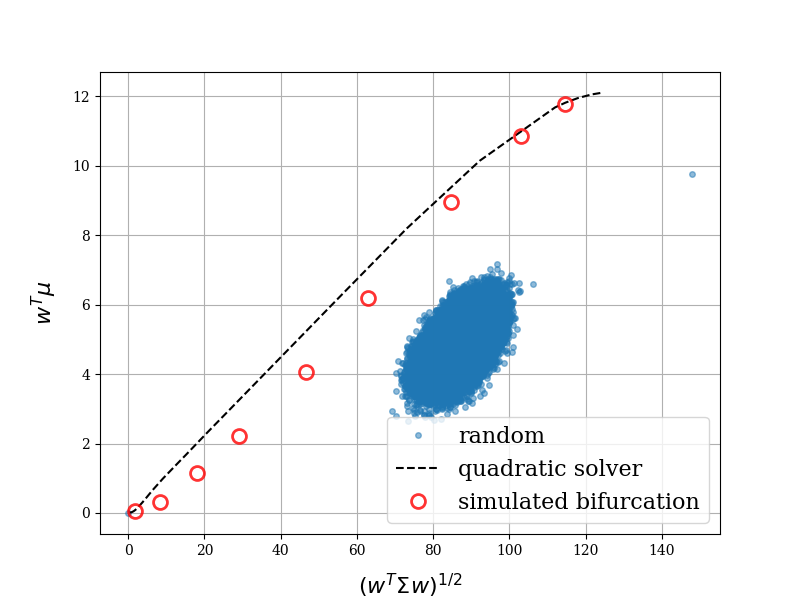}
 \caption{$N=100$, $\tilde{K}/N=15$ for various values of $\gamma$. Additionally $10^6$ random portfolios are shown. Calculation time for each point obtained by the SB-algorithm is around 0.4 seconds.}
\label{fig:N100K15}
\end{figure}

To illustrate this we perform the portfolio optimization at a fixed time point for 100 assets. If we limit the maximal number of units to be spent per asset to just $\tilde{K}/N=1$, the SB-algorithm finds the optimal solution as shown in fig. \ref{fig:N100K1} for a variety of different fixed parameters, in our example $\epsilon_p\in [0.001, 0.1]$ and $\Delta_t \in [0.01, 0.12]$ all generate the same results. The amount of different portfolios with similar values is comparably low due to the fact that only zero or one unit can be spent per asset. If we however increase the number of units to be spent to $\tilde{K}=1500$ where $\tilde{K}/N = 15$ units can be maximally spent per asset, i.e. 4 spins per asset instead of just one, the number of portfolios with almost identical value increases. In such scenarios we can detect deviations between the optimal solutions and those found by the SB-algorithm if no specific fine-tuning of the SB-parameters is conducted. This is illustrated in fig. \ref{fig:N100K15} where the SB-algorithm only finds close-to-optimal solutions with $\epsilon_p=0.05$ and $\Delta_t = 0.02$. This behaviour has a particular accent for $\gamma$ values that allow to balance risk and return, for extreme values of the risk aversion parameter however, this was not observed. In those limits also the number of portfolios with almost identical value decreases. Note that in both fig. \ref{fig:N100K1} and fig. \ref{fig:N100K15}, a set of random portfolios is plotted as well. This however is for illustrative purposes only, since the number of possible combinations is extremely large and random samples cannot help us detect the entire universe of possible portfolios. What is nevertheless remarkable in those two figures is the fact that without any fine-tuning the SB-algorithm finds close-to-optimal solutions in 0.4 seconds on a single CPU out of $2^{400}$ configurations and the global optimum even faster out of $2^{100}$ possibilities.

\begin{figure}[h]
 \includegraphics[width=0.5\textwidth, angle=0]{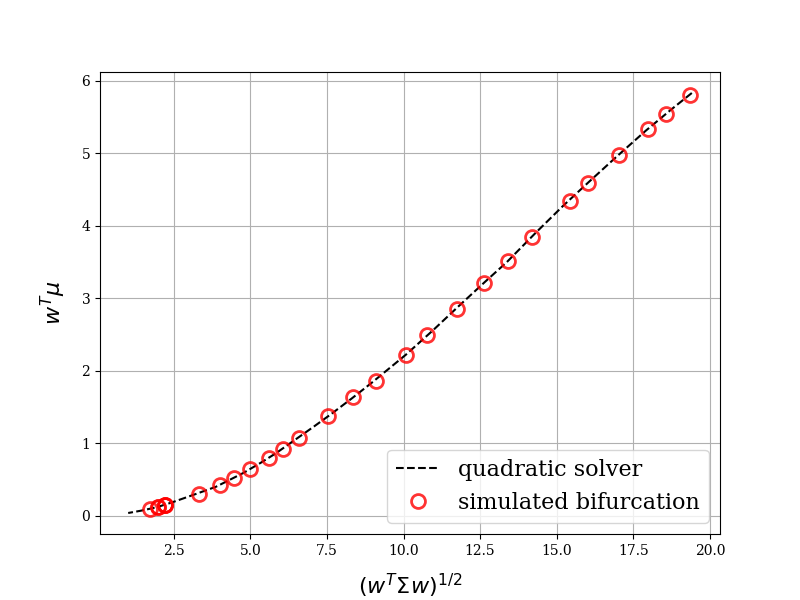}
 \caption{Portfolio with $N=1000$ assets, a total capital of $\tilde{K}=1000$ units, where $1$ can be spent maximally per asset, either purchase or not.  Computation time of the SB-approach is around 1 seconds per point, which can be pushed into the sub-second regime with the mentioned parameter tuning.}
\label{fig:verylarge1000_quadraticsolv}
\end{figure}

For smaller values of $\tilde{K}/N$, the parameter selection has proven to be much more forgiving. Even for large systems up to $1000$ assets the SB-algorithm successfully finds the optimal solutions. This is illustrated in fig. \ref{fig:verylarge1000_quadraticsolv}, where optimal portfolios with $N=1000$ are obtained regardless of the exact choice of SB-parameters in a certain interval, i.e. changing $\epsilon_p \in [0.01, 0.1]$ and $\Delta_t=[0.01,0.1]$ generates the same displayed solutions. This flexibility allows to chose aggressive SB-parameters, such that the computation of the optimal solution of systems of up to $N=1000$ and $\tilde{K}/N=1$ takes less than 1 second on a Desktop CPU.

In the previous section \ref{sec:opttradingtraj} we discussed the behaviour of systems in the limit of zero trading costs. This is a particularly useful limit since it allows us to test if the optimal trajectory coincides with the set of individually optimized portfolios at each point in time. Some trajectories found by the SB-agorithm however, have not shown the success observed in fig. \ref{fig:trajectory_numericalcheck} and without careful fine-tuning do not coincide exactly with the global optimum at $c=0$. For those cases this strongly indicates that the corresponding trajectories obtained for non-zero trading costs are also only close-to-optimal. First preliminary experiments suggest that this is not directly connected with the number of combinations, but again to the amount of trajectories with values very close to the optimum.

Note that this numerical cross-check with the trajectory at zero trading cost can also assist in tuning the SB-parameters. If the optimal solution is not known due to the high amount of possible configurations, a set of SB-parameters is selected that reproduces the reference configurations in the limit of vanishing trading costs. Note that the trading costs are elements of the interaction matrix in the Ising formulation, and hence this approach is assumed to work only for small enough values of $c$.

In this section we have identified that without proper fine-tuning of the SB-parameters we can get solutions that are only close to the optimum. This was however only observed for systems that show a large amount of portfolios or trajectories with values very close to the optimum. We have yet not established a proper dynamic parameter selection framework and resorted to manual adjustments in this initial phase of research. A mathematically rigorous investigation is necessary to understand the connection between the algorithm's parameter and the distribution of the different portfolio-values and the magnitude of their numerical differences. This is subject of ongoing investigations and will serve as the main key to construct a dynamic SB-parameter selection framework that allows for optimal portfolio optimization.

\section{Conclusion}
We have shown that the SB-algorithm can successfully be used to find optimal solutions for the integer portfolio problem as well as the integer trading trajectory problem.

We have investigated portfolios with up to 1000 assets and our first performance investigation on a single desktop CPU has already confirmed the power of the SB-algorithm for its novel use-case as a portfolio and trading trajectory optimizer. For the investigated portfolios the computation time does not exceed a couple of seconds and truly shows an unprecedented speed in finding optimal and close to optimal solutions. In a next step we will move the optimization framework to our GPU cluster for parallel computing and hope to share results in the near future that capture a substantial amount of the actively traded assets.

The presented formulation in the Ising representation incorporates the upper quantity limits as hard constraints and would also allow for individual asset dependent upper quantity constraints. With a proper penalty term in the Ising description we are able to control the costs of rebalancing the portfolio from one point in time to the next, which can also straightforwardly be made asset and time dependent.

For a large number of assets, in combination with a large number of units/capital to distribute per asset, the SB-algorithm has detected optimal but sometimes also only close-to-optimal solutions. For larger systems with a large amount of almost degenerate ground-states, the global optimum was often only detected after careful fine-tuning of the SB-parameters. Note however, that the rudimentary standard set of parameters that was successfully used across a wide range of problems already allowed to obtain solutions very close to the optimum. We have not established a framework which maps the problem settings to a suitable set of SB-parameters, that guarantee finding the ground-state. In this work we have resorted to manual adjustments of this important and delicate step, and consider a detailed parameter discussion the main key for further improvements in this direction.

Another future challenge is the introduction of further constraints, such as cardinality or fixed quantity constraints. The inclusion of further constraints can however quickly lead to the introduction of many ancillary spins and also has the potential to disrupt the evolution of the SB-algorithm into the ground-state. The constraints beyond the hard-wired limitations presented in this work are therefore considered one of the biggest challenges to make the SB-algorithm applicable for a variety of real world constraints in portfolio optimization problems.

The presented formulation allows to control also time and asset dependent trading activities, lacks however a proper translation from real word trading concepts to concrete numerical penalty values in the Ising formulation. There is therefore a strong interest to develop a more elaborate trading cost framework and to introduce rebalancing scenarios that are closer to real world applications. 

Despite the open challenges, this work has shown the first highly successful and incredibly fast portfolio optimization with the simulated bifurcation algorithm. An approach that we believe will see a wide range of applications in many other fields of finance as well.

\bibliographystyle{asmems4}


\end{document}